\documentclass[12pt,oneside]{article}
\usepackage[english]{babel}
\usepackage{graphics,graphicx,epsfig}
\makeindex
\newcommand{\be}{\begin{equation}}
\newcommand{\ee}{\end{equation}}
\newcommand{\bea}{\begin{eqnarray}}
\newcommand{\eea}{\end{eqnarray}}

\def\asec{$''$ cy$^{-1}$}

\def\bb{\bibitem}
\def\rfr#1{eq.(\ref{#1})}
\def\rfrs#1#2{eqs.(\ref{#1})-(\ref{#2})}
\def\Rfr#1{Eq.(\ref{#1})}

\def\eqi{\begin{equation}}
\def\eqf{\end{equation}}
\begin{document}  
\begin{titlepage} 
\begin{flushright}
\today\\
\end{flushright}
\vspace{.5cm}
\begin{center}
{\LARGE How the orbital period of a test particle is modified by the Dvali-Gabadadze-Porrati gravity?\\}
\vspace{0.5cm}
\quad\\
{Lorenzo Iorio\\
\vspace{0.5cm}
\quad\\
Viale Unit\`a di Italia 68, 70125,
Bari, Italy\\ \vspace{0.5cm}
}
\vspace*{0.5cm}

{\bf Abstract\\}
\end{center}

{\noindent \small In addition to the pericentre $\omega$, the mean anomaly $\mathcal{M}$ and, thus, the mean longitude $\lambda$, also the orbital period $P_b$ and the mean motion $n$ of a test particle are modified by the Dvali-Gabadadze-Porrati gravity. While the correction to $P_b$ depends on the mass of the central body and on the geometrical features of the orbital motion around it, the correction to $n$ is independent of them, up to terms of second order in the eccentricity $e$. The latter one amounts to about $2\times 10^{-3}$ arcseconds per century. The present-day accuracy in determining the mean motions of the inner planets of the Solar System from radar ranging and differential Very Long Baseline Interferometry ($\Delta$VLBI) is $10^{-2}-5\times 10^{-3}$ arcseconds per century, but it should be improved in the near future when the data from the spacecraft to Mercury and Venus will be available. } \end{titlepage}
\newpage
\setcounter{page}{1}
\vspace{0.2cm}
\baselineskip 14pt

\setcounter{footnote}{0}
\setlength{\baselineskip}{1.5\baselineskip}
\renewcommand{\theequation}{\mbox{$\arabic{equation}$}}
\noindent

\section{Introduction}
According to the Dvali-Gabadadze-Porrati (DGP) model of gravity \cite{DGP00}, our Universe is a (3+1) space-time brane embedded in a five-dimensional Minkowskian bulk with an extra-spatial dimension which is flat and infinite. Many important consequences of cosmological interest, mainly related to the observed cosmic acceleration, can be traced out from such a scenario \cite{Lue05}.  

A very appealing feature of DGP gravity is that global information about the the current cosmological phase can be obtained from local observational tests which could be conducted at Solar System scales. Lue and Starkman \cite{LS03} derived a secular  extra-pericentre advance $\dot\omega$ for test particles moving around a central body of mass $M$ in almost circular orbits. Remarkably, such an effect is independent both of the particular features of the orbital path (up to second order terms in the eccentricity $e$) and of the central body; it amounts to about $5\times 10^{-4}$ arcseconds per century (\asec). It lies at the edge of the present-day planetary ephemeris accuracy and the conditions for its possible detection have been preliminarily investigated in \cite{LS03, DGZ03} and, with some  more details, in \cite{Ior05a}. The full set of DGP orbital perturbations, for non-circular orbits, have been derived in \cite{Ior05b}. It turns out that the first nonvanishing terms depending on the eccentricity are of second order and are too small to be detected. Moreover,  the DGP gravity also affects the mean anomaly $\mathcal{M}$ with an extra-secular advance so that the planetary mean longitudes $\lambda$ precess at a $\sim 10^{-3}$ \asec\ rate.
In \cite{Ior05c} it has been proposed to explain  the recently observed secular increase of the Astronomical Unit \cite{KraBru04, Sta05} in terms of DGP gravity.

In this paper we further investigate the impact of the DGP model on the orbital motion of test bodies by working out its effect on the period of revolution. The obtained results are compared with the present-day observational accuracy.

\section{The impact of DGP gravity on the orbital revolution}
\subsection{The anomalsitic period and the mean motion}
One of the six Keplerian orbital elements in terms of which it is possible to parameterize the orbital motion of a planet is the mean anomaly  $\mathcal{M}$. It is defined as $\mathcal{M}\equiv n(t-T_0)$ where $n$ is the mean motion and $T_0$ is the time of pericentre passage. The mean motion $n\equiv 2\pi/ P_b$, in turn, is inversely proportional to the time elapsed between two consecutive crossings of the pericentre, i.e. the anomalistic period $P_b$. 
In Newtonian mechanics $n=\sqrt{GM/a^3}\equiv\bar{n} $, where $a$ is the semi-major axis of the orbit; moreover, the apsidal line is a fixed direction in space\footnote{We assume that the central body which acts as source of the gravitational field is spherically symmetric.}, so that the anomalistic period is equal to the sidereal period, i.e.  the time elapsed between two consecutive crossings of a fixed direction with respect to the distant quasars. If the pericentre suffers small advances after every orbital revolution, as it happens in the DGP model, the anomalistic period is not equal to the sidereal period. We will focus on the anomalistic period because it is easier to determine in ordinary planetray data reduction processes.

\subsection{The DGP correction to the anomalistic period and the mean motion}
In the DGP model a free-crossover parameter $r_0$ is present: it is fixed to $r_0\sim 5$ Gpc by the latest Type IA 
Supernov\ae\ measurements \cite{LS03}. Beyond it, gravity gets substantially modified with cosmological implications, while at scales much smaller than $r_0$ the usual Newton-Einstein gravity is restored, apart from small but relevant modifications. In such a regime the DGP gravity induces an extra radial acceleration  \cite{Gruz05, LS03, Ior05b}
\eqi A_{\rm DGP}=\mp\left(\frac{c}{2r_0}\right)\sqrt{ \frac{GM}{r} }\equiv \frac{\kappa}{\sqrt{r}},\label{acc}  \eqf which can be considered as a small perturbation to the larger Newtonian monopole. The upper sign refers to the standard FLRW cosmological phase, while the lower sign is related to the self-accelerating cosmological phase.

Let us investigate the impact of \rfr{acc} on the anomalistic period of a test particle. To this aim, we must consider 
the Gauss equation for the mean anomaly. In the case of a purely radial perturbing acceleration $A$, it reduces to
\eqi\frac{d\mathcal{M}}{dt}=n-\frac{2}{na}A\left(\frac{r}{a}\right)+\frac{(1-e^2)}{nae}A\cos f,\label{gauss} \eqf
where  $f$ is the true anomaly reckoned from the pericentre.
By inserting \rfr{acc} in \rfr{gauss} it is possible to obtain
\eqi\left.\frac{d\mathcal{M}}{dt}\right|_{\rm DGP}=n\left\lbrace \ 1-\frac{\kappa}{n^2 a}\left[\frac{2\sqrt{r}}{a} 
-\frac{(1-e^2)}{e}\frac{\cos f}{\sqrt{r}}\right] \right\rbrace.\label{gauss2}  \eqf
\Rfr{gauss2} must be evaluated on the unperturbed Keplerian ellipse characterized by
\begin{eqnarray}
r&=&\frac{a(1-e^2)}{1+e\cos f},\label{uno} \\
\frac{d\mathcal{M}}{df}&=&\left(\frac{r}{a}\right)^2\frac{1}{\sqrt{1-e^2}}\label{due},\\
n&=&\bar{n} \label{oo}.
\end{eqnarray}  

By inserting \rfrs{uno}{oo} in \rfr{gauss2}  and expanding $(1+e\cos f)^{-5/2}$ and $(1+e\cos f)^{-3/2}$ to first order in $e$, one gets
\eqi dt\sim \frac{1}{ \bar{n}}\left\lbrace \frac{(1-e^2)^{3/2}}{(1+e\cos f)^2} +\frac{\kappa(1-e^2)^2}{ {\bar{n}}^2 a^{3/2} }\left[2-\left(5e+\frac{1}{e}\right)\cos f+\frac{3}{2}\cos^2 f\right]\right\rbrace df.\label{pio} \eqf
The pericentre-to-pericentre time can, thus, be obtained by integrating \rfr{pio}  from 0 to $2\pi$
\eqi P_b=\frac{2\pi}{\bar{n}}\left[1+\frac{11}{4}\frac{\kappa(1-e^2)^2}{{\bar{n}}^2 a^{3/2}}\right].\label{roar} \eqf
The DGP correction to the Keplerian period is
\eqi P_b^{\rm (DGP)}=\mp\frac{11\pi}{8}\left(\frac{c}{r_0}\right)\frac{a^3(1-e^2)^2}{GM}.\label{PDGP} \eqf
It is interesting to note that \rfr{PDGP} depends on the characteristics of both the central body and of the orbital path around it.

Since $P_b$ is determined via the mean motion (see Section \ref{measu}), it is useful to define it from \rfr{roar}   as
\eqi n\equiv\frac{2\pi}{P_b}=\frac{\bar{n}}{1+\frac{11}{4}\frac{\kappa(1-e^2)^2}{{\bar{n}}^2 a^{3/2}}}\equiv \bar{n}+\Delta n^{\rm (DGP)}.\eqf
It turns out that the DGP correction to the Keplerian mean motion
\eqi \Delta n^{\rm (DGP)}\sim\pm\frac{11}{8}\left(\frac{c}{r_0}\right)(1-e^2)^2\eqf is independent of the features of both the central body and of the test particle's orbit (to order $\mathcal{O}(e^2)$).
\subsection{A theoretical caveat}
Before discussing the comparison with the present-day observations, the following theoretical reamarks are in order.

\Rfr{acc} comes from the correction to the Newtonian potential of a Schwarzschild source found in \cite{Gruz05, LS03}.
Such a potential is obtained within a certain approximation
which is valid below the Vainshtein scale\footnote{For a Sun-like star $r_{\star}$ amounts to about 100 parsec.} $r_{\star}=(r_g r_0^2)^{1/3}$, where $r_g=2GM/c^2$ is the Schwarzschild radius of the central object. However, it is not yet clear, at present,  whether this potential can match continuously onto a four-dimensional Newtonian potential above the Vainshtein scale, and then, also match onto the five-dimensional potential above the crossover scale\footnote{A very similar problem exists in massive gravity, where it was shown by numerical and analytical methods that the matching between such approximate solutions fails, always requiring a naked singularity.} $r_0$. An alternative solution that smoothly interpolates between the different regions was discussed in \cite{GaIg05a,GaIg05b}. The correction to the Newtonian potential arising from that solution below the Vainshtein scale is somewhat different from what used here. In particular, it is reduced by a multiplicative factor smaller than unity.
As a consequence, the predictions are also different. 

Should the approximate solution used in the present work will  turn out to be the wrong one, this fact must be accounted for in the confrontation with the observational data.
\section{The present-day observational accuracy}\label{measu} 
The orbital period  of a planet is not directly measurable but it is obtained by the mean motion which is one of the many parameters which can be determined by fitting the observations in the standard ephemeris generation process. The DGP term could, then, be determined by including a correction $\Delta n$ to the standard reference values of the mean motions in the set of parameters to be fitted when the whole set of available observational data are processed against a converged ephemeris, as recently done, e.g., by Pitjeva in \cite{Pit05} for the extra-advances of perihelia.

According to Standish \cite{Sta05}, the accuracy with which it is possible to determine the mean motions of the inner planets of the Solar System from radar ranging\footnote{Among the various quantities directly observable (ranges, range-rates, angles, etc.), the ranges are the most accurately determined ones: the present-day residuals for the inner planets amount to about 1 km.} and $\Delta$VLBI amounts to  $10^{-2}-5\times 10^{-3}$ \asec: the magnitude of $\Delta n_{\rm DGP}$ is one order of magnitude smaller amounting to $2\times 10^{-3}(1-e^2)^2$ \asec.
However, ranging data from the inner planets should become much more accurate when the radar waves will be reflected back by the future, forthcoming spacecraft (Messenger and BepiColombo to Mercury and Venus Express to Venus) instead of the badly known surfaces of the planets themselves.
In regard to Mars, its orbital accuracy is corrupted at a rate of a few km/decade by the perturbations induced by the asteroids whose masses are poorly known \cite{Sta05}.

As it can be inferred from \rfr{PDGP}, the outer planets of the Solar System are more sensitive to the DGP modifications of gravity: e.g. the corrections to the orbital periods for them range from $5\times 10^{-2}$ s (Jupiter) to $3\times 10^{1}$  s (Pluto). However, it must be considered that for them mainly optical observations are available: their accuracy  cannot compete with radar-ranging. Moreover, their orbital periods amount to tens or hundreds of years.

\section{Conclusions}
In this paper we have worked out the effect of the the Dvali-Gabadadze-Porrati multidimensional gravity model on the anomalistic period and mean motion of a test particle. It turns out that, while the correction to the orbital period depends both on the mass of the central body which acts as source of the gravitational field and on the shape and the size of the test particle's orbit, the correction to the mean motion is independent of them, up to terms of second order  in the eccentricity. The correction to the mean motion amounts to about $2\times 10^{-3}$ arcseconds per century. The present-day accuracy in determining the mean motions of the inner planets of the Solar System from radar ranging and differential Very Long Baseline Interferometry is $10^{-2}-5\times 10^{-3}$ arcseconds per century, but it should be improved in the near future when the data from the spacecraft Venus Express, BepiColombo and Messenger to Venus and Mercury will be available.
\section*{Acknowledgements}
I thank E.M. Standish (JPL) for useful correspondence.

\end{document}